# Multi-wavelength Spin Dynamics of Defects in Hexagonal Boron Nitride

Ivan Zhigulin[†,1,2], Nicholas P. Sloane[†,*,1,2], Benjamin Whitefield[1,2], Jean-Philippe Tetienne[3], Mehran Kianinia[*,1,2], and Igor Aharonovich[1,2]

[1] School of Mathematical and Physical Sciences, University of Technology Sydney, Ultimo, New South Wales 2007, Australia
[2] ARC Centre of Excellence for Transformative Meta-Optical Systems, University of Technology Sydney, Ultimo, New South Wales 2007, Australia
[3] Department of Physics, School of Science, RMIT University, Melbourne, VIC 3001, Australia

† These authors contributed equally to this work.
* To whom correspondence should be addressed: N.P.S Nicholas.sloane@uts.edu.au, M.K. Mehran.Kianinia@uts.edu.au

## Abstract

*Optically addressable solid-state spin defects are essential platforms for quantum sensing and information processing. Recently, single spin defects with combined $S = 1$ and $S = ½$ spin transitions were discovered in hexagonal boron nitride (hBN). In this work we unveil their excitation dynamics. In particular, we study the effects of the excitation wavelength on the spin-dependent fluorescence and the spin dynamics of these peculiar quantum spin defects. We find that changing the excitation wavelength leads to a threefold enhancement in both the optically detected magnetic resonance (ODMR) contrast and the corresponding magnetic field sensitivity. In addition, we find that the excitation wavelength has a strong impact on the photodynamics of spin complex emitters. Our work presents valuable insights to the mechanistic understanding of spin complex emitters in hBN and highlights the importance of excitation wavelength for optimising their performance in quantum sensing and quantum technologies.*

## Introduction

Optically addressable spin systems have emerged as promising platforms for quantum communication and sensing[1–3]. Defects in three-dimensional host materials such as diamond and silicon carbide have been extensively studied for quantum sensing, demonstrating sensitivity to magnetic fields, electric fields, and temperature[4]. Alongside these conventional platforms, hexagonal boron nitride (hBN) has emerged as a complimentary host for optically addressable spin defects, with its van der Waals structure providing unique advantages[5–13]. Notably, the two-dimensional nature of hBN allows atomically thin layers to be exfoliated directly onto target samples[14–18] or seamlessly integrated into photonic structures and optoelectronic devices[19–21].

Currently, the most studied spin defect in hBN is the boron vacancy ($V_B^-$) which has a spin-triplet (S = 1) ground state[22–26]. This defect exhibits optically detected magnetic resonance (ODMR) at zero field and at room temperature. It has already been utilised for sensing magnetic fields, temperature, pressure, and strain[27–30]. However, so far single $V_B^-$ defects have not been identified, which limits its broad adaptation for quantum information. In addition, their low quantum efficiency and the dark zero phonon line has further motivated a search for bright narrowband single spin defects in hBN.

Recently, a new class of spin-active defects has been discovered in hBN and other solid-state materials, referred to as the spin complex[10–12,31–33]. These spin complex defects are brighter, show a clear single photon emission, and have zero-phonon lines (ZPLs) spanning the range of visible to near-infrared wavelengths (500-750 nm)[12]. Most intriguingly, the spin complex in hBN exhibits both S = 1 and S = ½-like spin transitions in ODMR spectra. The S = 1 transitions are attributed to an electron pair occupying a strongly coupled triplet state, whereas the S = ½ transitions correspond to a delocalised, weakly coupled electron spin pair[10,12]. The coexistence of these spin manifolds provides an exciting new platform for quantum sensing and spin manipulation in hBN.

In this work, we study the spin-photon dynamics of the spin complex in hBN, focusing specifically on the excitation wavelength and its effect on the spin dynamics. We find that the ODMR contrast and the stability of the photoluminescence (PL) of the spin complex are strongly influenced by the excitation wavelength, with ODMR transitions tripling in contrast and reaching nearly 100 %.

**Results and Discussion**

The spin complex in hBN is schematically illustrated in Figure 1a which exhibits two distinct spin manifolds, a localised strongly coupled spin pair (orange) and a delocalised weakly coupled spin pair (blue). The energetic structure formed by the spin complex is shown in Figure 1b and includes the optically active S = 0 ground and excited states (|g⟩ and |e⟩, respectively), as well as the triplet states in the metastable (MS) regime which consists of two distinct spin manifolds. The first manifold has both electrons occupying the same defect resulting in a localised, strongly coupled spin pair with the well defined spin states (S = 1). These energetic levels are shown in the red box in Figure 1b as |+1⟩, |0⟩, and |-1⟩ corresponding to $m_s$ = +1, 0, and -1, respectively. Following charge transfer of one electron to a nearby defect, the spin pair becomes delocalised, resulting in weaker coupling and transition into the second manifold (S = {1,0}). In this configuration the spin states are instead effectively characterised by a state of mixed singlet and triplet character ($|ST_0⟩$) alongside states with pure triplet character ($|T_±⟩$) as shown in Figure 1b. From the weakly coupled manifold the electron can reform back to the optically active defect via charge transfer, relaxing to the S = 0 ground state |g⟩, from which it can be re-excited and undergo fluorescence.

As $|ST_0⟩$ and $|T_±⟩$ decay at different rates to |g⟩, applying microwaves resonant with the splitting between the weakly coupled spin states leads to a change in the fluorescence, resulting in a detectable ODMR signal. Similarly, before charge transfer to the weakly coupled manifold occurs, application of resonant microwaves with the splitting between the

strongly coupled spin states modifies the population of |ST$_0$⟩ and |T$_\pm$⟩, resulting in an ODMR signal. Previous studies have shown that no ODMR is observed at zero field, implying that the strongly coupled spin-pair population cascades through the weakly coupled manifold prior to relaxation to the ground state[10,12]. At low fields, the energetic splitting between the spin states in the weakly coupled manifold is nearly degenerate leading to rapid spin mixing and loss of spin information.

Figure 1c shows the typical continuous wave (CW)-ODMR spectrum from a spin complex emitter with signatures of transitions between both, the strongly coupled (S = 1) states and the weakly coupled (S = {1,0}) states. Resonances in the ODMR spectrum are assigned to the different transitions based on solutions to the spin Hamiltonian incorporating the S = 1 and S = {1,0} states (see Supplementary Information Section 1). As no ODMR is detected at zero field for the spin complex emitter, the spectrum was acquired with an applied out-of-plane magnetic field of ~50 mT. The magnetic field strength was kept approximately the same for all measurements throughout this work. The transitions within the strongly coupled manifold, namely -1 ↔ 0, 0 ↔ +1, and the -1 ↔ +1 resonances can be seen at approximately 0.7, 2.1, and 2.9 GHz, respectively. The peak at ~1.1 GHz corresponds to the second harmonic of the 0 ↔ +1 transition. Furthermore, we note that the −1 ↔ +1 transition at 2.9 GHz indicates a double-quantum transition ($\Delta m_s = 2$), which is forbidden under standard spin-selection rules. Nevertheless, this transition produces a strong ODMR signal and has been widely reported for spin complex emitters. The transition within the weakly coupled manifold (S = {1,0}) is observed at 1.4 GHz and is commonly referred to as the -½ ↔ +½ transition, as its response is reminiscent to the behaviour of a single spin-½ particle.

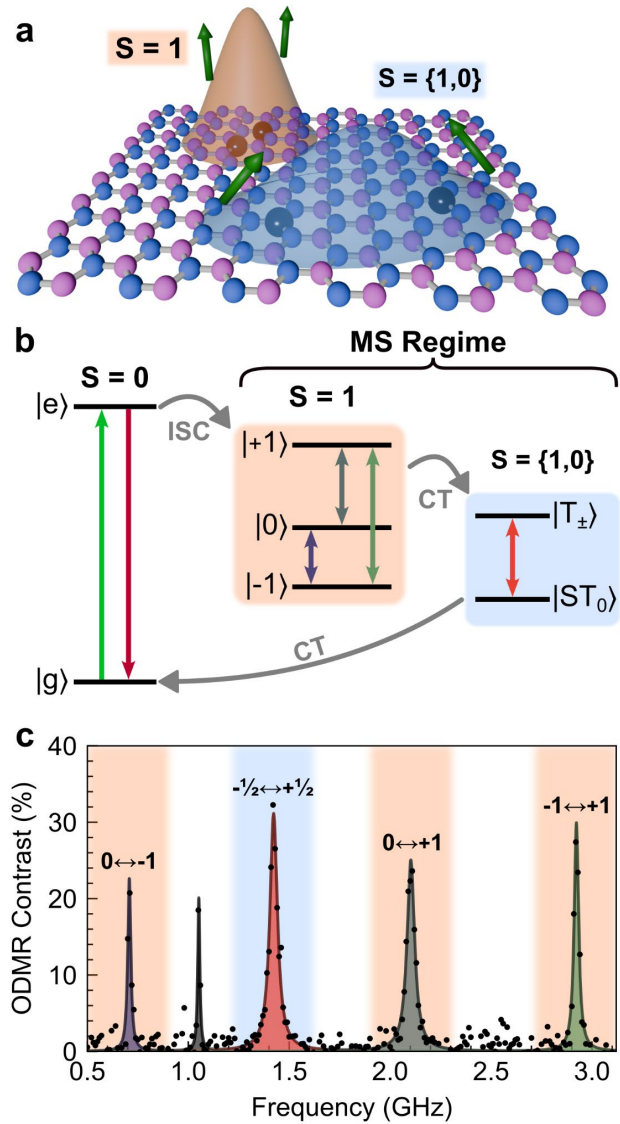

*Figure 1: The spin complex in hBN. (a) Illustration of the spin complex wavefunctions in hBN represented by a localised, strongly coupled spin pair (orange), and a delocalised weakly coupled spin pair (blue). (b) Energy level diagram of the spin complex consisting of the singlet ground and excited state, |g⟩ and |e⟩, and the metastable (MS) regime under an applied external field. Within the MS regime, the strongly coupled spin states |+1⟩, |0⟩, and |-1⟩ correspond to $m_s$ = +1, 0, and -1 respectively. Following charge transfer (CT), the spin pair becomes weakly coupled, leading to the formation of states with mixed singlet–triplet character (|$ST_0$⟩) and states with pure triplet character (|$T_\pm$⟩). (c) CW-ODMR spectrum of a spin complex emitter showing the -1 ↔ +0, -½ ↔ +½, 0 ↔ +1, and the -1 ↔ +1 resonances at ~50 mT.*

Prior to entering the MS regime via intersystem crossing (ISC), the spin complex defect must first be driven to the excited state by an absorption of incoming photons. To locate emitters, scanning confocal microscopy was employed. Figure SI2a shows a region of hBN flake with multiple optically active emitters, highlighting the emitter of interest. PL spectra of this emitter under 532 nm and 633 nm excitations are presented in Figure SI2b, with no discernible differences observed between the two wavelengths. Similarly, Figure S3a,b shows the autocorrelation measurements of the emitter, which exhibit comparable antibunching behaviour ($g^{(2)}(0) < 0.5$) under both 532 nm and 633 nm excitations. We note that this emitter could not be resolved under 405 nm excitation, likely due to a reduced absorption cross section at this wavelength or excitation into an optically inactive state.

In stark contrast to the PL signatures of this emitter under different excitation wavelengths, the spin-dependent fluorescence is strongly impacted by the excitation wavelength. Figure 2a presents CW-ODMR spectra of the -½ ↔ +½ and 0 ↔ +1 transitions under 532 nm and 633 nm excitations. Remarkably, the ODMR contrast is approximately three times greater when excited with 633 nm relative to 532 nm excitation, increasing from 36% to 98% for the -½ ↔ +½ transition. This behaviour is similarly observed for all ODMR transitions outlined in Figure 1b, with the changes in the -1 ↔ +0 and -1 ↔ +1 transitions shown in Figure SI4.

Figure 2b shows the relationship between the measured ODMR contrast for the -½ ↔ +½ and 0 ↔ +1 transitions and the excitation power ($P_{exc}$), normalised by the saturation power ($P_{sat}$) extracted from Figure SI5 for both excitation wavelengths. The extracted saturation powers from Figure SI5 were found to be 151 μW and 79 μW for 532 nm and 633 nm excitations, respectively. Note the rise and subsequent fall in contrast as a function of optical excitation power, which was also observed previously for the spin complex in hBN[10]. Yet, the contrast under 633 nm excitation consistently remains higher than that under 532 nm excitation across all excitation powers, suggesting that this behaviour is not described by the optical excitation power. Instead, the enhanced contrast may potentially arise from different couplings between the excited states accessed by each wavelength and the spin-active MS states. We measured two additional spin complex emitters and observed similar excitation wavelength-dependence of the ODMR signal (see Supplementary Information Section 3). We note however, that this effect appears to be emitter-dependent, with different excitation wavelengths producing varying contrast strengths across different emitters.

To investigate the wavelength-dependent photodynamics of the emitter, PL counts were continuously measured over an extended time interval. Figure 2c,d shows trajectories of the PL over time of the emitter excited with 532 nm and 633 nm, respectively. For additional comparison, PL count histograms for both excitation wavelengths are also included in Figure 2c,d. Under 532 nm excitation, PL of the emitter remains relatively stable for the duration of the measurement. In contrast, under 633 nm excitation the emitter exhibits pronounced "blinking" behaviour, intermittently switching between bright emission and non-luminescent intervals as seen in the variation in the histogram in Figure 2d. To further probe the non-trivial photodynamics, second-order autocorrelation with longer timescales (~1 ms) were performed under both 532 nm and 633 nm excitations. These measurements investigate photon bunching that may account for the observed blinking behaviour, shown in Figure 2e. Both 532 nm and 633 nm excitations showed bunching signatures, with significantly stronger bunching for the former wavelength. This is expected as longer timescales of the blinking behaviour contribute to the photon statistics, introducing additional decay lifetimes in the autocorrelation histogram. Besides stronger coupling to the MS regime, the effect could also be due to the presence of optically inactive states that are also being populated, including other excited energy charge states or electron transfer to a nearby charge trap.

Based on the ODMR contrast results in Figure 2c,d and the photodynamics observed in Figure 2c–e, we propose a conceptual model in Figure 2f that explains the wavelength-dependent behaviour within the current understanding of the energetic structure of the spin complex outlined in Figure 1b. Different excitation wavelengths, corresponding to different photon energies, selectively populate distinct excited states within the singlet (optically active, $S = 0$) regime. In Figure 2f, this is represented by 633 nm photons promoting population into the excited state $|e_1\rangle$ and 532 nm light populating $|e_2\rangle$, where each state corresponds to different vibronic levels or distinct energy levels. The relative energetic positions of these excited states with respect to the $S = 1$ triplet states in the MS regime determine the efficiency of ISC. In the proposed model in Figure 2f, $|e_1\rangle$ is predicted to couple strongly to the MS regime, accounting for the enhanced ODMR contrast and potentially the blinking behaviour as the states in the MS regime are optically inactive. In contrast, $|e_2\rangle$ is predicted to have weaker coupling to the MS regime, resulting in less population transfer to the triplet states, thus decreasing ODMR contrast and reducing non-luminescent times. We applied an existing rate model, used to describe spin complex emitters in hBN, to the power dependent contrast in Figure 2b[13], and observed a close fit to the recorded data. The model assumes a higher ISC rate from $|e_1\rangle$ to the MS regime as compared to $|e_2\rangle$. Additionally, the polarisation rate of the different triplet states following ISC is dependent on the optical pumping power, resulting in the decrease in contrast at higher laser powers as initially explained by Dréau *et al.* for NV⁻ centers in diamond[34]. Details on the derivation and explanation of the model can be found in Supplementary Information Section 4.

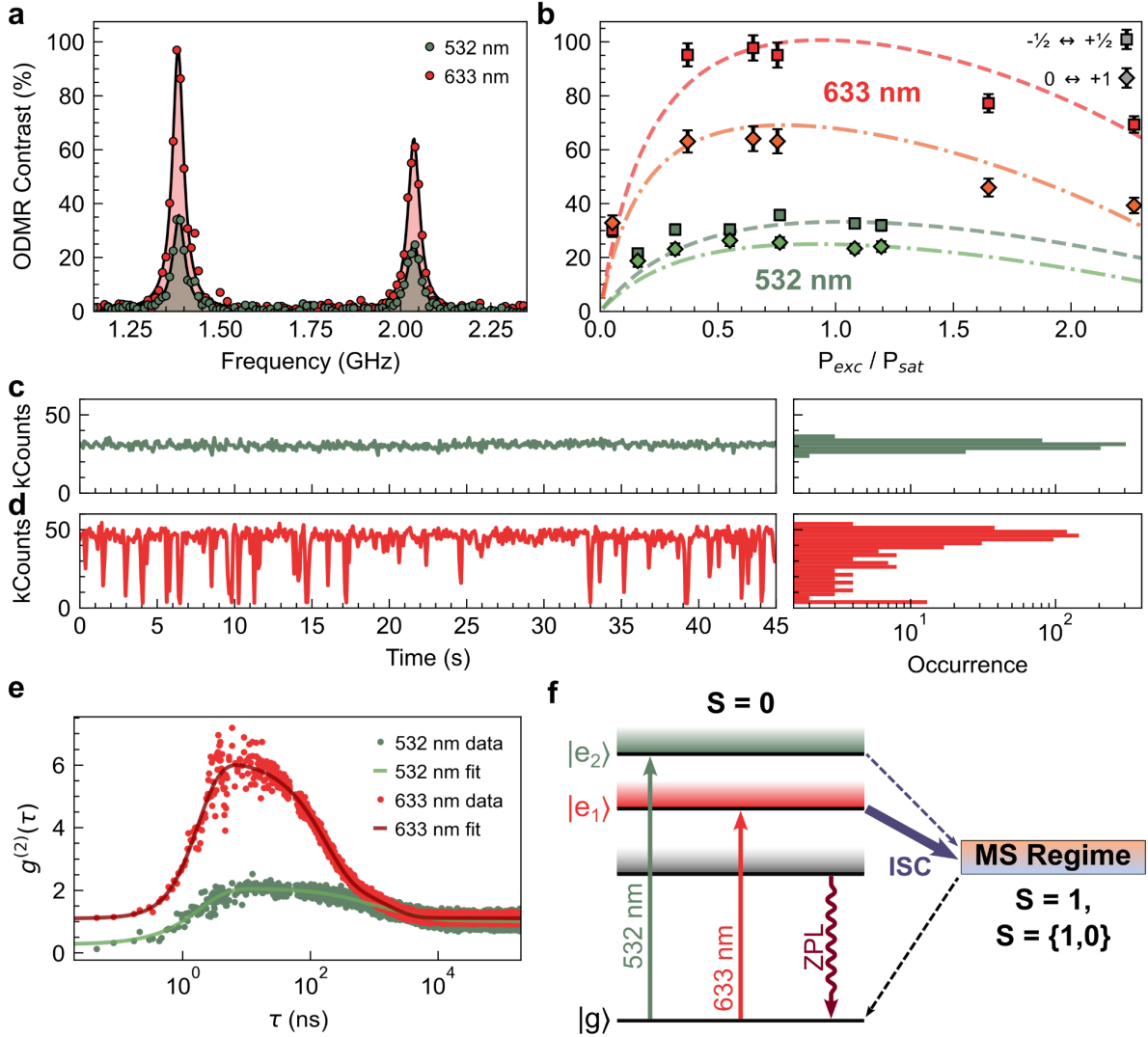

*Figure 2: Behaviour of the spin complex under different excitation wavelengths. (a) CW-ODMR spectra of the emitter excited with 532 nm (green) and 633 nm (red) showing the relative contrast for the -½ ↔ +½, and 0 ↔ +1 transitions. (b) Measured ODMR contrast of the -½ ↔ +½ (squares) and 0 ↔ +1 (diamonds) resonances under both 532 and 633 nm excitations plotted as a function of optical power normalised by the saturation power ($P_{exc}/P_{sat}$). The dashed lines represent the modelled contrast as a function of $P_{exc}/P_{sat}$, an explanation of the model can be found in Supplementary Section 4. (c), (d) Time traces of the emitter fluorescence under 532 nm and 633 nm excitation, respectively, with the corresponding photon count histograms (1 kCount bin width) shown on the right for each excitation wavelength. (e) Long-timescale second-order autocorrelation measurements, recorded under 532 nm and 633 nm excitation at optical powers of 63 µW and 50 µW, respectively. (f) Energy level diagram illustrating a potential mechanism for population of the MS regime under different excitation wavelengths. Coupling between the state populated by 633 nm excitation and the MS regime is anticipated to be stronger than that of the state populated by 532 nm excitation, leading to increased ODMR contrast and partially responsible for the observed blinking behaviour.*

To further investigate the excitation wavelength-dependence of the spin complex behaviour discussed above, we extended the same confocal microscope configuration to conduct simultaneous co-excitation with two wavelengths (532 nm and 633 nm). The optical layout is summarised in Figure 3a as a simplified schematic, and is discussed in detail in methods.

Using the co-excitation scheme, we first examine PL dynamics in the time domain. Figure 3b shows PL count traces recorded at a fixed 633 nm excitation power of 80 µW while varying the 532 nm power from 0 to 150 µW. As discussed in the previous section, excitation with 633 nm alone produces pronounced blinking. However, introducing 532 nm excitation progressively suppresses this blinking, with higher 532 nm powers leading to increasingly stable emission. This trend is more clearly visualised in the extracted count histograms shown in Figure 3c, where distributions are plotted on a logarithmic occurrence scale for better visibility.

In addition to suppressed blinking, introduction of 532 nm excitation initially leads to an increase in total PL intensity, followed by counts saturation at higher powers. This is shown in Figure 3d, where the extracted total PL intensities are plotted as a function of 532 nm power. The dependence can be expressed as $I(P)$, which follows a power saturation model of the form:

$$I(P) = I_{Offset} + \frac{I_\infty P}{P + P_{Sat}} \quad (1)$$

The relationship accounts for the constant PL contribution ($I_{Offset}$) from the fixed 633 nm excitation power of 80 µW and provides luminescence counts $I_\infty$ at saturation power $P_{Sat}$ of 8.33 µW.

From the histogram plots in Figure 3c it is clearly shown that excitation with 633 nm alone results in two distinct count distributions. Using their minimum overlapping point, we define the threshold value (black dashed line), which separates the histograms into "dark" (below threshold) and "bright" (above threshold) states. This can be clearly visualised using a simple illustration schematic in the inset of Figure 3e, where two Gaussian profiles represent two emission states divided by a threshold. The threshold value is scaled as counts intensity ratio for each 532 nm power obtained from Figure 3d data and follows the power saturation model from above. Using this approach, the density of dark state counts can be consistently extracted at different co-excitation powers of 532 nm (shown in Figure 3e). Under sole 633 nm excitation, approximately 11% of the counts are in the dark state, whereas addition of only 2 µW of 532 nm excitation reduces this fraction to ~4%. Increasing the 532 nm power further progressively suppresses the dark state. Equal contributions of 532 nm and 633 nm and dominant 532 nm excitation (≥80 µW) lead to a single bright state above the threshold, as shown in Figure 3c and inset of Figure 3e.

The observed behaviour suggests that under 633 nm excitation the spin complex more frequently cycles through the MS regime which involves charge transfer between different spin manifolds, thus increasing the blinking rate. Addition of 532 nm excitation may enable optical repumping from the S = 1 state into a higher energy excited level[35]. Phonon-assisted decay then allows the electron to relax into a ZPL emissive state that has a reduced coupling to the MS regime. This results in stabilised emission with lower blinking at the expense of ODMR contrast, as electrons are removed from the MS regime.

This interpretation is consistent with the ODMR measurements shown in Figure 3f, where 633 nm-only excitation yields the highest contrast. Adding low 532 nm power (5 µW) reduces the contrast by ~4% for -½ ↔ +½ and ~2% for 0 ↔ +1 transitions. At the highest 532 nm power, the contrast decreases to ~30% for both transitions. Stepwise measurements as a function of 532 nm power in Figure 3g reveal a gradual reduction in contrast. However, the different rates at which the dark state population and ODMR contrast decrease suggest that repumping alone may not fully account for the observations. Analysis of other transitions of the S = 1 manifold is in Supplementary Section 5.

There are two additional factors contributing to the different trends of the dark state population and ODMR contrast. First, under sole 532 nm excitation the emitter becomes optically active at powers of ≳15 µW. Thus, during co-excitation in the high power regime, increasing dominance of 532 nm excitation competes with the 633 nm pathway, reducing the overall coupling to the MS regime. Second, the presence of an alternative non-luminescent decay pathway cannot be excluded. Indeed, it is common for hBN emitters to display blinking characteristics due to having additional charge states[36,37] or charge traps in the vicinity that are activated through different laser excitation[38]. Combination of these factors contribute to the changes of populations of dark/bright states and ODMR contrast.

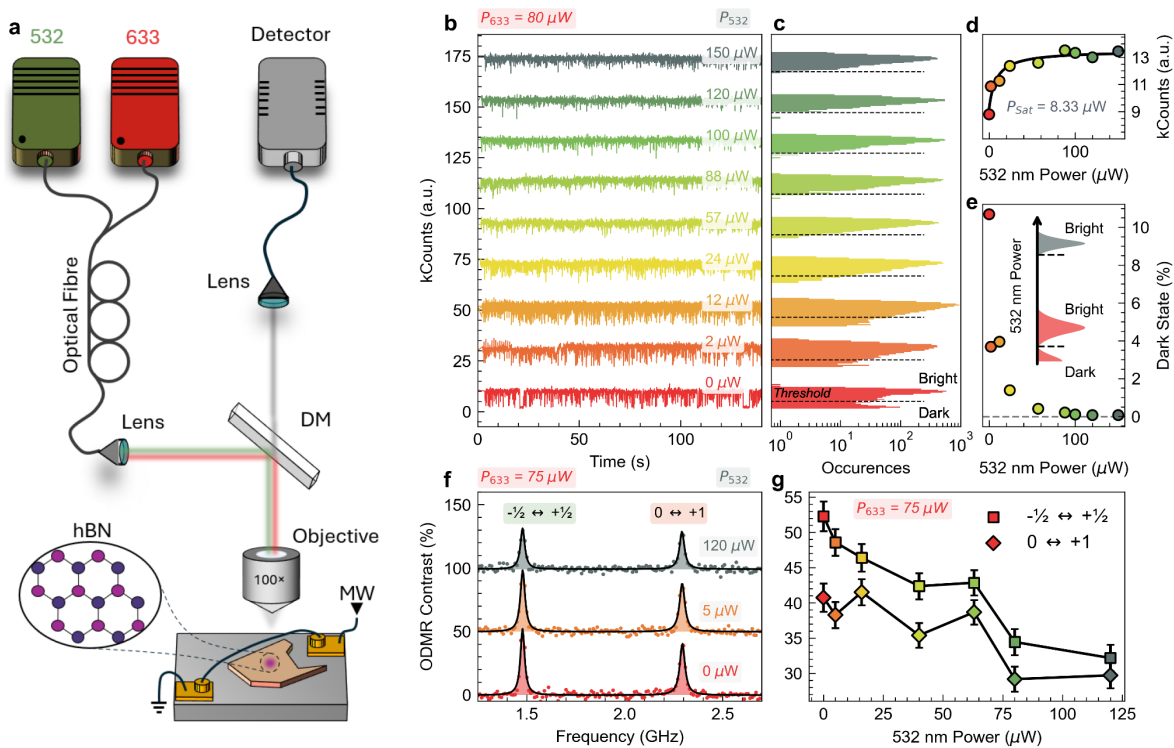

*Figure 3: Co-excitation of the spin complex with 633 nm and 532 nm. (a) Schematic of the setup used for the co-excitation measurement. 532 nm and 633 nm lasers were coupled into a 1×2 single-mode optical fibre that reflected light off a dichroic mirror (DM) into an objective. PL was collected in reflection using the same objective and coupled into multimode fibre. Microwaves (MW) were delivered onto the sample using a suspended copper wire. (b) Time traces of the emitter fluorescence under constant 633 nm power of 80 µW and varying powers of 532 nm (annotated in the figure). (c) Corresponding photon count histograms for each 532*

*nm power. Black dashed lines indicate a threshold that distinguishes between bright and dark states.* **(d)** *PL counts saturation as a function of varying 532 nm power and a fixed 633 nm power of 80 μW.* **(e)** *Percentage of counts found below the threshold (in the dark state) as a function of varying 532 nm laser power. Inset: diagram showing distribution of counts at 0 μW and 150 μW of 532 nm excitation power, where the former has two distinct peaks corresponding to dark and bright states, while the latter has no dark state.* **(f)** *CW-ODMR spectra of the emitter under co-excitation using 75 μW of 633 nm and three selected powers of 532 nm (0 μW, 5 μW, and 120 μW). It displays reductions in contrast for the for the -½ ↔ +½ and 0 ↔ +1 transitions.* **(g)** *ODMR contrast as a function of 532 nm power for the -½ ↔ +½ (squares) and 0 ↔ +1 (diamonds) transitions.*

In the final section, we probe the spin-complex under multicolour excitation at cryogenic temperatures. Performing these measurements at 25 K allows cleaner spectral resolution by suppressing thermal broadening and reducing background contributions present at room temperature. Figure 4a shows CW-ODMR spectra acquired under 532 nm, 610 nm, and 650 nm excitations. Using 532 nm, ODMR contrast of the -½ ↔ +½ and 0 ↔ +1 transitions remains below 35 %, consistent with room temperature data. Changing the excitation to 610 nm, results in a pronounced increase in ODMR contrast, with both resonances rising above 50 %. Changing excitation wavelength even closer to the ZPL (650 nm) yields similar contrast levels, remaining above 50 %. The -1 ↔ +1 transition displayed similar behaviour and is shown in Figure SI11. These results highlight once more that for this emitter, excitation wavelengths below ~600 nm are less effective at populating the states in the MS regime responsible for high ODMR contrast. Throughout these measurements Figure 4b confirms that spectral characteristics of the emitter remained unchanged. This then opens avenues for conducting PL spectral response to examine how microwave driven spin transitions affect individual emission signatures.

Under 532 nm excitation, the PL spectrum exhibits two dominant peaks separated by ~10 meV, consistent with coupling to a longitudinal acoustic phonon mode of the sideband[39]. Figure SI12 shows that both spectral components display minimal fluctuations in intensity and spectral wandering (remaining below ~0.3 nm), indicating stable emission from both peaks. Following, we probe the effect of microwave driven spin transitions on individual spectral components. The applied microwave frequency was set either on resonance with the -½ ↔ +½ or 0 ↔ +1 spin transitions, or to an off-resonant frequency for comparison.

Changes in the relative peak intensities were then analysed. Representative spectra acquired under identical optical conditions with the microwave field tuned on- and off-resonance are shown in Figure 4c. Double-Lorentzian fits to these spectra yield nearly identical area ratios for the two peaks across all microwave frequencies. This result is summarised in Figure 4d, where the extracted peak areas are plotted as histograms and their ratios are annotated. The ratios deviate by less than 0.02, indicating that both spectral components couple evenly with the MS regime. This suggests that the electron relaxation pathways have equal ISC probability to the S = 1 spin manifold. Thus, the demonstrated technique reliably probes spin dynamics via spectral signatures. We consequently apply this analysis at room temperature to assess the influence of stronger phonon contributions. Similarly, the results revealed minimal variations in peak-area ratios between on- and off-resonant microwave excitation. The corresponding analysis is provided in Supplementary Information Section 7.

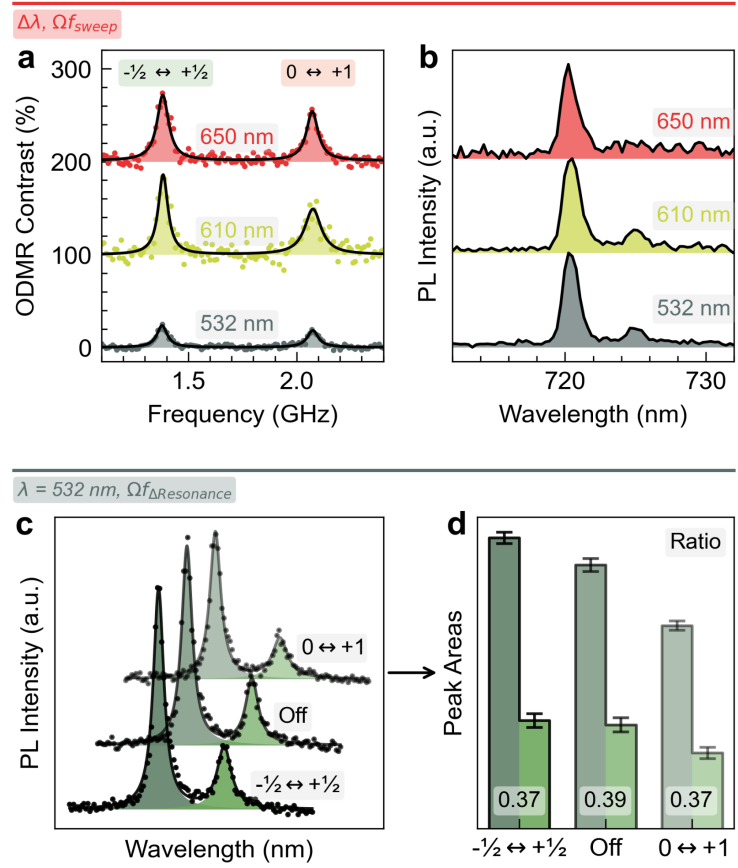

*Figure 4: Multicolour excitation and PL-resolved ODMR at cryogenic temperature (25 K). Excitation wavelength and microwave frequency are abbreviated as λ and Ωf, respectively. (a) CW-ODMR spectra under 532 nm, 610 nm, and 650 nm excitations. (b) PL spectra of the emitter at 532 nm, 610 nm, and 650 nm excitations. (c) Fitted PL spectra of the emitter with applied microwaves tuned to resonance with the -½ ↔ +½ and 0 ↔ +1 transitions, alongside microwaves detuned to an off-resonant frequency. (d) Histograms of areas of each peak for the corresponding microwave transitions extracted from data in (c). Bar ratios for each transition are annotated in the plot.*

The results presented in this work are especially relevant for the development of quantum sensing based on spin complex defects. In particular the dependence of ODMR contrast on excitation wavelength has direct implications for magnetic field sensing with the spin complex, as the DC magnetic field sensitivity ($\eta_{DC}$) is calculated as[34,40]:

$$\eta_{DC} = \mathcal{P}_L \frac{h}{g_e \mu_B} \frac{\Delta\nu}{C\sqrt{R}} \quad (2)$$

Where $\mathcal{P}_L$ is a constant based on the Lorentzian profile of the peak (≈0.77), h is Planck's constant, $g_e$ = 2 is the electron g-factor, $\mu_B$ is the Bohr magneton, $\Delta\nu$ is the ODMR linewidth, C is the contrast, and R is the photon count rate. We find that for the −½ ↔ +½ transition, $\eta_{DC}$ improves from 23.1 ± 2.2 µT/$\sqrt{Hz}$ under 532 nm excitation to 7.9 ± 0.7 µT/$\sqrt{Hz}$ under 633 nm excitation. Similarly, the 0 ↔ +1 transition shows sensitivity improvement from 33.1 ± 4.3 µT/$\sqrt{Hz}$ to 12.4 ± 1.5 µT/$\sqrt{Hz}$ under 532 nm and 633 nm excitations, respectively. This establishes the excitation wavelength optimisation as a practical pathway for significantly improving contrast readout of the spin complex, useful in quantum sensing applications and applicable to other quantum systems[41].

Additionally, the wavelength-dependent photodynamics of the spin complex could enhance super-resolution imaging via ground-state depletion microscopy (GSD), as previously demonstrated for hBN emitters[35]. In typical GSD microscopy, doughnut-shaped beam profiles are used to drive emitters surrounding the central intensity null into dark states, thereby increasing the effective resolution at the centre of the null[42]. As such, excitation wavelengths that more effectively promote ISC into metastable states are anticipated to enhance the performance of GSD microscopy. This significantly exceeds resolution achieved by a standard confocal microscope that is limited by light diffraction. Coupled with the spin-dependent fluorescence of the spin complex allows super-resolution imaging to realise quantum sensing at few nanometre resolution using spatial maps of both PL intensity and ODMR contrast.

**Conclusion**

In summary, we performed detailed spin and photodynamics studies of the spin complex emitters in hBN. Remarkably upon using 633 nm excitation wavelengths, ODMR contrast (and correspondingly the sensitivity) was found to be three times greater than that of the same emitter excited with 532 nm laser. Additionally, the emitter showed different photodynamics, evidenced by strong blinking behaviour and photon bunching present in the autocorrelation measurements. Based on this behaviour, we proposed a conceptual model in which different excitation wavelengths populate distinct excited states with varied coupling strengths to the optically inactive states in the metastable regime. From this, we employed a co-excitation scheme to stabilise PL of the emitter, using relatively weak 532 nm excitation powers in combination with constant 633 nm excitation. Finally, we investigated behaviour of the emitter at cryogenic temperatures to gain further insight into the spin-dependent PL of the spin complex through spectrally resolved ODMR. Our findings highlight the critical role of excitation wavelength in controlling both the spin-dependent behaviour and photodynamics of spin complex emitters, providing new strategies for optimising their performance in quantum sensing and photonic applications.

**Methods**

*Optical Characterisation*

All optical measurements were carried out using home-built scanning confocal microscope systems operating in reflection measurement. Emitters were addressed using 532 nm and 633 nm lasers coupled into the same 1×2 single-mode optical fibre (Thorlabs) and out-coupled through a fibre launcher equipped with a double-achromatic broadband coated lens. This ensured an identical Gaussian beam profile, diameter, and spatial mode for both excitation wavelengths. Downstream optical components included a red dichroic mirror (Semrock) and an achromatic 100× objective (Mitutoyo Plan Apo), typically high numerical aperture (≥0.7) to spatially resolve individual emitters. The collected PL signal was coupled into a multimode fibre and detected using either an avalanche photodiode detector (Excelitas APD) or spectrometer charge coupled detector imaging (Andor SR303I-Newton CCD) using 300 l/mm grating. Correlation measurements used a multimode 50:50 beamsplitter fiber with coincidence counts analysed by a counter module (Swabian Instruments Timetagger 20).

Room temperature measurements were carried out in ambient conditions. For cryogenic measurements samples were cooled down to 25 K using a closed-loop helium cryostat (Attocube attoDRY 800) and kept at pressure of ~$10^{-5}$ mbarr.

*Optically Detected Magnetic Resonance*

For ODMR measurements a static magnetic field was applied using a permanent NdFeB magnet (grade N35) positioned perpendicular to the sample surface to lift the spin-state degeneracy. Microwave excitation was provided by a radio-frequency signal generator (AnaPico APSIN 4010), with the output delivered to the sample via a suspended copper wire placed in close proximity to the flake. The microwave signal was amplified using a high-power microwave amplifier (Mini-Circuits ZHL-16W-43-S+). The microwave frequency was swept within the 0.5~4.0 GHz resonance range while monitoring the photoluminescence signal to obtain ODMR spectra.

**Data Availability**

The data supporting the findings of this study are available within the paper and its supporting information files.

**Acknowledgements**

The authors acknowledge financial support from the Australian Research Council (CE200100010, FT220100053, and DP240103127), the Air Force Office of Scientific Research (FA2386-25-1-4044).

**References**

1. Aharonovich, I., Englund, D. & Toth, M. Solid-state single-photon emitters. *Nat. Photonics* **10**, 631–641 (2016).
2. Awschalom, D. D., Hanson, R., Wrachtrup, J. & Zhou, B. B. Quantum technologies with optically interfaced solid-state spins. *Nat. Photonics* **12**, 516–527 (2018).
3. Wolfowicz, G. *et al.* Quantum guidelines for solid-state spin defects. *Nat. Rev. Mater.* **6**, 906–925 (2021).
4. Schirhagl, R., Chang, K., Loretz, M. & Degen, C. L. Nitrogen-Vacancy Centers in Diamond: Nanoscale Sensors for Physics and Biology. *Annu. Rev. Phys. Chem.* **65**, 83–105 (2014).

5. Stern, H. L. *et al.* Room-temperature optically detected magnetic resonance of single defects in hexagonal boron nitride. *Nat. Commun.* **13**, 618 (2022).
6. Scholten, S. C. *et al.* Multi-species optically addressable spin defects in a van der Waals material. *Nat. Commun.* **15**, 6727 (2024).
7. Stern, H. L. *et al.* A quantum coherent spin in hexagonal boron nitride at ambient conditions. *Nat. Mater.* **23**, 1379–1385 (2024).
8. Singh, P. *et al.* Violet to Near-Infrared Optical Addressing of Spin Pairs in Hexagonal Boron Nitride. *Adv. Mater.* **37**, 2414846 (2025).
9. Gilardoni, C. M. *et al.* A single spin in hexagonal boron nitride for vectorial quantum magnetometry. *Nat. Commun.* **16**, 4947 (2025).
10. Gao, X. *et al.* Single nuclear spin detection and control in a van der Waals material. *Nature* **643**, 943–949 (2025).
11. Robertson, I. O. *et al.* A charge transfer mechanism for optically addressable solid-state spin pairs. *Nat. Phys.* 1–7 (2025) doi:10.1038/s41567-025-03091-5.
12. Whitefield, B. *et al.* Narrowband quantum emitters in hexagonal boron nitride with optically addressable spins. *Nat. Mater.* 1–8 (2026) doi:10.1038/s41563-025-02458-6.
13. Whitefield, B. *et al.* Photodynamics and Temperature Dependence of Single Spin Defects in Hexagonal Boron Nitride. Preprint at https://doi.org/10.48550/arXiv.2512.07067 (2025).
14. Durand, A. *et al.* Optically Active Spin Defects in Few-Layer Thick Hexagonal Boron Nitride. *Phys. Rev. Lett.* **131**, 116902 (2023).
15. Huang, M. *et al.* Wide field imaging of van der Waals ferromagnet Fe3GeTe2 by spin defects in hexagonal boron nitride. *Nat. Commun.* **13**, 5369 (2022).
16. Healey, A. J. *et al.* Quantum microscopy with van der Waals heterostructures. *Nat. Phys.* **19**, 87–91 (2023).
17. Robertson, I. O. *et al.* Detection of Paramagnetic Spins with an Ultrathin van der Waals Quantum Sensor. *ACS Nano* **17**, 13408–13417 (2023).
18. Gao, X. *et al.* Nanotube spin defects for omnidirectional magnetic field sensing. *Nat. Commun.* **15**, 7697 (2024).
19. Caldwell, J. D. *et al.* Photonics with hexagonal boron nitride. *Nat. Rev. Mater.* **4**, 552–567 (2019).
20. Moon, S. *et al.* Hexagonal Boron Nitride for Next-Generation Photonics and Electronics. *Adv. Mater.* **35**, 2204161 (2023).
21. Loh, L., Wang, J., Grzeszczyk, M., Koperski, M. & Eda, G. Towards quantum light-emitting devices based on van der Waals materials. *Nat. Rev. Electr. Eng.* **1**, 815–829 (2024).
22. Gottscholl, A. *et al.* Initialization and read-out of intrinsic spin defects in a van der Waals crystal at room temperature. *Nat. Mater.* **19**, 540–545 (2020).
23. Gottscholl, A. *et al.* Room temperature coherent control of spin defects in hexagonal boron nitride. *Sci. Adv.* **7**, (2021).
24. Mu, Z. *et al.* Excited-State Optically Detected Magnetic Resonance of Spin Defects in Hexagonal Boron Nitride. *Phys. Rev. Lett.* **128**, 216402 (2022).
25. Qian, C. *et al.* Unveiling the Zero-Phonon Line of the Boron Vacancy Center by Cavity-Enhanced Emission. *Nano Lett.* **22**, 5137–5142 (2022).

26. Gong, R. *et al.* Coherent dynamics of strongly interacting electronic spin defects in hexagonal boron nitride. *Nat. Commun.* **14**, 3299 (2023).
27. Gottscholl, A. *et al.* Spin defects in hBN as promising temperature, pressure and magnetic field quantum sensors. *Nat. Commun.* **12**, 4480 (2021).
28. Lyu, X. *et al.* Strain Quantum Sensing with Spin Defects in Hexagonal Boron Nitride. *Nano Lett.* **22**, 6553–6559 (2022).
29. Yang, T. *et al.* Spin defects in hexagonal boron nitride for strain sensing on nanopillar arrays. *Nanoscale* **14**, 5239–5244 (2022).
30. Whitefield, B., Toth, M., Aharonovich, I., Tetienne, J.-P. & Kianinia, M. Magnetic Field Sensitivity Optimization of Negatively Charged Boron Vacancy Defects in hBN. *Adv. Quantum Technol.* **8**, 2300118 (2025).
31. Luo, J., Geng, Y., Rana, F. & Fuchs, G. D. Room temperature optically detected magnetic resonance of single spins in GaN. *Nat. Mater.* **23**, 512–518 (2024).
32. Vaidya, S. *et al.* Coherent Spins in van der Waals Semiconductor GeS2 at Ambient Conditions. *Nano Lett.* **25**, 14356–14362 (2025).
33. Liu, W. *et al.* Experimental Observation of Spin Defects in the van der Waals Material GeS2. *Nano Lett.* **25**, 16330–16339 (2025).
34. Dréau, A. *et al.* Avoiding power broadening in optically detected magnetic resonance of single NV defects for enhanced dc magnetic field sensitivity. *Phys. Rev. B* **84**, 195204 (2011).
35. Kianinia, M. *et al.* All-optical control and super-resolution imaging of quantum emitters in layered materials. *Nat. Commun.* **9**, 874 (2018).
36. Stern, H. L. *et al.* Spectrally Resolved Photodynamics of Individual Emitters in Large-Area Monolayers of Hexagonal Boron Nitride. *ACS Nano* **13**, 4538–4547 (2019).
37. Boll, M. K., Radko, I. P., Huck, A. & Andersen, U. L. Photophysics of quantum emitters in hexagonal boron-nitride nano-flakes. *Opt. Express* **28**, 7475–7487 (2020).
38. Comtet, J. *et al.* Wide-Field Spectral Super-Resolution Mapping of Optically Active Defects in Hexagonal Boron Nitride. *Nano Lett.* **19**, 2516–2523 (2019).
39. Preuss, J. A. *et al.* Resonant and phonon-assisted ultrafast coherent control of a single hBN color center. *Optica* **9**, 522–531 (2022).
40. Gao, X. *et al.* High-Contrast Plasmonic-Enhanced Shallow Spin Defects in Hexagonal Boron Nitride for Quantum Sensing. *Nano Lett.* **21**, 7708–7714 (2021).
41. Todenhagen, L. M. & Brandt, M. S. Optical and electrical readout of diamond NV centers in dependence of the excitation wavelength. *Appl. Phys. Lett.* **126**, 194003 (2025).
42. Hell, S. W. & Kroug, M. Ground-state-depletion fluorescence microscopy: A concept for breaking the diffraction resolution limit. *Appl. Phys. B* **60**, 495–497 (1995). 995).